\documentclass[11pt]{article}
\usepackage{rotating}

\usepackage{lscape}

\usepackage{graphicx}
\begin{document}
\title{\bf Masses of RR Lyrae Stars with Different Chemical Abundances
in the Galactic Field}

\author{{V.\,A.~Marsakov, M.\,L.~Gozha, V.\,V.~Koval'}\\
{Southern Federal University, Rostov-on-Don, Russia}\\
{e-mail:  marsakov@sfedu.ru, gozha\_marina@mail.ru, vvkoval@sfedu.ru}}
\date{accepted \ 2019, Astronomy Reports, Vol. 63, No. 3, pp. 203-211}

\maketitle

\begin {abstract}

The surface gravities and effective temperatures have 
been added to a compilative catalog
published earlier, which includes the relative abundances of 
several chemical elements for 100 field
RR Lyrae stars. These atmoshperic parameters and evolutionary 
tracks from the Dartmouth database are
used to determine the masses of the stars and perform a 
comparative analysis of the properties of RR Lyrae
stars with different chemical compositions. The masses of 
metal-rich ($\rm{[Fe/H]}> -0.5$) RR Lyrae stars with
thin disk kinematics are in the range $(0.51-0.60)M_{\odot}$. 
Only stars with initial masses exceeding $1M_{\odot}$ 
can reach the horizontal branch during
the life time of this subsystem. To become an RR Lyrae variable,
a star must have lost approximately half of its mass during the 
red-giant phase. The appearance of such young, metal-rich RR Lyrae 
stars is possibly due to high initial helium abundances of their 
progenitors. According to the Dartmouth evolutionary tracks for 
Y = 0.4, a star with an initial mass as low as $0.8M_{\odot}$ 
could evolve to become an RR~Lyrae 
variable during this time. Such stars should have lost 
$(0.2-0.3)M_{\odot}$ in the red-giant phase, which seems quite 
realistic. Populations of red giants and RR Lyrae stars with
such high helium abundances have already been discovered in the 
bulge; some of these could easily be transported to the solar 
neighborhood as a consequence of perturbations due to 
inhomogeneities of the Galaxy’s gravitational potential.

\end{abstract}


\maketitle

\section{Introduction}

This paper continues our studies of RR Lyrae variables
in the Galactic field initiated in \cite{1,2,3}. These
papers describe a catalog we have compiled containing
the positions, velocities, and metal abundances
of 415 field RR Lyrae stars, along with the relative
abundances [el/Fe] of 12 elements for 100 RR Lyrae
stars, including four $\alpha$-process elements (Mg, Ca,
Si, and Ti). We have used this catalog to study the
relationships between the chemical properties of field
RR Lyrae stars and their spatial and kinematic characteristics.
In particular, we demonstrated that, despite
the high ages usually claimed for these stars, the
RR Lyrae population includes representatives of the
youngest Galaxy subsystem, the thin disk. We also
pointed out the problem of the existence of metal-rich
RR Lyrae stars with $\rm{[Fe/H]}> -0.5$. According to theoretical
computations, the initial masses of these stars
must have had in order for them to reach the horizontal
branch are fairly low, $(0.55 - 0.8) M_{\odot}$ \cite{4}, and the
evolution of such stars should take more than 10 billion
years; i.\,e., more than the age of the thin-disk
subsystem in the Galaxy. Higher-mass metal-rich
stars evolve to the region of the red clump, outside
the instability strip, and cannot become variables.
However, the kinematics and chemical abundances of
these stars derived in our study testify that they very
likely belong to the thin disk and have younger ages.
Some 70 years ago, Kukarkin \cite{5} noted the presence
of ``peculiar'' field RR Lyrae stars (called short-period
Cepheids at that time) with periods below
0.43 days, which were not present in globular clusters
and showed a stronger concentration to the Galactic
plane than RR Lyrae variables with longer periods.
Later, most of these were found to be rich in metals.
Despite the long history of their studies, the nature of
metal-rich RR Lyrae stars remains incompletely understood.
A semi-empirical explanation of the origin
of metal-rich, comparatively young RR Lyrae stars
suggested in \cite{6} is that they result from the loss of a
considerable fraction of a star's mass ($\sim 0.5M_{\odot}$) during
the red-giant evolution phase. This idea did not
become popular, although it was also never disproved.
The physical processes of metal-rich and metal-poor
RR Lyrae stars were also found to differ. For example,
the processes operating during the pulsations of these
variables were studied in detail in \cite{7}, with considerable
differences in the envelope kinematics found for
metal-rich ($\rm{[Fe/H]}> -1.0$) and metal-poor RR Lyrae
stars. Chadid et\,al. \cite{7} concluded that, although all
RRab variables were horizontal-branch stars, metal-rich
RR Lyrae stars had a different specific nature of
their own.

It was suggested in \cite{8} that the metal-richest
and longest-period RR Lyrae variable in our list
(KP Cyg) is most likely a classical Cepheid with an
ultra-short period. It is possible that all or some
metal-rich RR Lyrae stars could also be Cepheids
pulsating in overtones with periods below one day.
Such Cepheids have already been discovered in the
Large and Small Magellanic Clouds through the
OGLE project. Therefore, we suggested in \cite{1,3} that
we should look for the origins of the kinematic and
chemical youth of metal-rich RR Lyrae stars in their
classification as variable stars. In this case, however,
their masses should be even higher than for lower-temperature
RR Lyrae stars, in contradiction with
the general trend for the masses of horizontal-branch
star to increase with decreasing temperature. Indeed,
the effective temperatures and surface gravities were
found in \cite{7} to be higher for metal-rich RR Lyrae stars
compared to metal-poor ones. At the same time,
according to theoretical computations, the masses
of horizontal-branch stars should increase with their
surface gravity (log\,$g$); however, on the contrary, their
masses decrease with increasing metallicity. Thus, it
is difficult to draw any conclusions about the mass of
a particular RR Lyrae star \textsl{a priori}.

Taking all this into account, our study is aimed at
determining the masses of field RR Lyrae stars with
published elemental abundances and atmospheric
parameters using theoretical evolutionary tracks,
together with a comparative analysis of the properties
of metal-rich RR Lyrae stars in order to improve our
understanding of their nature.


\section {INPUT DATA}

We found the effective temperatures $T_{\textrm{eff}}$ and surface
gravities log\,$g$ for all 100 RR Lyrae variables in
our catalog in the same 25 publications from 1995--2017 
that had earlier been our sources of information
on the elemental abundances in the atmospheres of
these stars \cite{1}. References to these publications can
be found in our online catalog.

The atmospheric parameters of periodically pulsating
RR Lyrae variables depend on their pulsation
phase. It is believed that the most accurate atmospheric
parameters can be obtained during the phase
of minimum brightness ($\phi \sim 0.8$), when the star's
atmosphere is essentially not affected by pulsations
(e.g., \cite{9,10}). The basic parameters of the stellar
atmospheres, $T_{\textrm{eff}}$ and log\,$g$, can be determining using
both spectroscopic and photometric methods. Some
researchers have recommended preferred parameter
values among those they have obtained using various
methods. When recommendation is expressed by the
authors of the studies used, we calculated the parameters
averaged over phase and over values obtained
using different methods. Taking into account only
values for phases close to the minimum brightness
did not lead to any significant change in the average
values for $T_{\textrm{eff}}$ and log\,$g$.

Most researchers had estimated the uncertainties
of the parameters they derived. Spectroscopic determinations
of $T_{\textrm{eff}}$ and log\,$g$ have uncertainties from
40 to 300 K and from 0.1 to 0.5 dex, with the mean
values being $\varepsilon(T_{\textrm{eff}}) = 115$ K 
and $\varepsilon$(log\,$g$) = 0.24. The
uncertainties of photometrically determined values of
$T_{\textrm{eff}}$ and log\,$g$ vary from 100 to 250 K and from 0.1
to 0.3 dex; the mean uncertainties quoted for this
method are $\varepsilon(T_{\textrm{eff}}) = 170$ K 
and $\varepsilon$(log\,$g$) = 0.18. The
modes of the uncertainties using any methods are
about 200 K for the effective temperature and 0.3 dex
for the surface gravity. If there was no information
available about the uncertainties in measured parameters,
we adopted these uncertainties.

The effective temperatures and surface gravities
for 40 RR Lyrae variables were determined in two
or more studies. For these stars, we calculated the
weighted mean parameters using weights that were
inversely proportional to the quoted uncertainties. We
used these same RR Lyrae stars to plot distributions
of the deviations of both parameters from the corresponding
weighted mean values for all the determinations.
These distributions were successfully represented
with Gaussian curves, indicating the random
character of the differences between parameters
obtained in different studies. The mean dispersions
provide estimates of the external agreement of the parameters
derived in different studies, 
$\varepsilon(T_{\textrm{eff}}) = 226 \pm 15$ K and 
$\varepsilon$(log\,$g$)$ = 0.25 \pm 0.02$. We can see that this
external agreement is comparable to the uncertainties
averaged over all the methods.

Table 1 presents atmospheric parameters for the
100 RR Lyrae stars with known elemental abundances.
The columns of this table contain (1) the
name of the star; (2) the metallicities from \cite{1}; (3) the
relative abundances of $\alpha$-process elements, [$\alpha$/Fe]
(see below); (4)--(6) data on the effective temperature:
the value adopted in this paper (TP) and the highest
and lowest $T_{\textrm{eff}}$ values among those recommended in
publications or averaged over phase and methods,
when $T_{\textrm{eff}}$ values were taken from two or more papers;
(7)--(9) analogous parameters for the surface gravity
log\,$g$; and (10) mass of the star determined using
evolutionary tracks (see below).

Figure 1a plots the surface gravity log\,$g$ versus the
effective temperature log\,$T_{\textrm{eff}}$ for the 100 RR Lyrae
stars of our sample. The filled black circles show
the metal-richest stars, filled gray circles those
with metallicities $-0.5 > \rm{[Fe/H]} > -1.0$, and open
triangles the lowest-metallicity stars. The temperature
and surface gravity uncertainties described
above result in uncertainties in the derived masses
$\varepsilon(M/M_{\odot}) \approx 0.015$. Figure~1 shows that the positions
of metal-rich RR Lyrae stars in this diagram
somewhat contradicts the conclusion of \cite{7}
that metal-rich RR Lyrae stars mainly have higher
values of log\,$g$ and of log\,$T_{\textrm{eff}}$ than do metal-poorer
stars: in contrast to \cite{7}, our sample contains an
appreciable number of low-metallicity RR Lyrae stars
with higher temperatures than those of metal-rich
stars. However, most of the low-metallicity RR Lyrae
stars that form a clump in the diagram are cooler.
Note that all the metal-poor RR Lyrae stars in the
clump (with the exception of HH Pup) belong to the
halo or thick disk according to their kinematics; i.\,e.,
they have a high ages that would enable initially low-mass
stars to reach the horizontal branch.

We used the spatial and kinematic data of Dambis
et al. \cite{11} in our sample of RR Lyrae stars \cite{1}, where
the magnitude in the $K_{S}$ infrared filter was applied
to introduce interstellar reddening corrections and to
calibrate the distances to the RR Lyrae stars. We
emphasize that RR Lyrae variables remain one of the
few kinds of objects that are easy to identify at considerable
distances, where direct parallax measurements
are not possible, even using modern instruments. It
is thus interesting to check how the absolute magnitudes
derived from the calibration agree with the
theoretical values. After replacing the metallicities
in the formula from \cite{11} with our spectroscopic data,
we calculated the relationship between the absolute
magnitudes and the variability periods of RR Lyrae
stars from the same catalog: 
$M_{Ks} = -0.769 + 0.088\cdot\rm{[Fe/H]}_{Sp} - 2.33\cdot log P_{F}$, 
where $\rm{[Fe/H]}_{Sp}$ are
the spectroscopic metallicities and $P_{F}$ are the fundamental
pulsation periods of the RR Lyrae stars.

Figure 1b plots the absolute magnitude $M_{Ks}$
versus the effective temperature for the same stars.
To facilitate a comparison of these two diagrams, we
plotted evolutionary tracks for stars of various masses
with the same input parameters on both diagrams.
In general, the stars with different metallicities have
approximately similar distributions, but the surface
gravities and calculated absolute magnitudes for
some stars contradict each other. In particular,
SDSS J1707+58 and KP Cyg have high log\,$g$ values,
but high luminosities. On the contrary, HH Pup has
a fairly low log\,$g$ value, but its luminosity is also low.
We will not consider the origins of such discrepancies
for each RR Lyrae star, as this requires a special
investigation. The stars displaying the strongest
deviations are marked in the diagrams. Note that
we were able to find only upper or lower limits to
the masses of FV Aqr, DO Vir, and CM Leo, and it
was not possible to determine the masses of CU Com
and SDSS J1707+58 because their log\,$g$ values were
outside the range specified by the limits of the grid of
models used for horizontal-branch stars.

Let us note some general patterns shown by
Fig.~1. It is striking that stars in the 
log\,$T_{\textrm{eff}}$ -- $M_{Ks}$
diagram are aligned almost parallel to the horizontal
axis, while they lie parallel to the theoretical lines for
constant masses in the 
log\,$T_{\textrm{eff}}$ -- log\,$g$ diagram. Two
of the highest-metallicity RR Lyrae stars (KP Cyg
and UY CrB) are the brightest in the log $T_{eff}$ -- $M_{Ks}$
diagram, although the other metal-rich RR Lyrae
stars have the lowest luminosities. Lower-metallicity
RR Lyrae stars ($-1.0 < \rm{[Fe/H]} < -0.5$) also lie
near the lower boundary of the diagram, while they
are considerably higher than the metal-richest stars
in the log\,$T_{\textrm{eff}}$ -- log\,$g$ diagram, and low-metallicty
RR Lyrae stars are only slightly brighter. Apparently
the atmospheric parameters determined spectroscopically
are more correct than the distances and absolute
magnitudes derived from calibration relations. In
addition, these parameters are also computed more
reliably using model atmospheres. Thus, we prefer to
determine the masses using these parameters.

\section {MASSES OF THE RR LYRAE STARS} 

We determined the masses of the RR Lyrae stars in
our sample using evolutionary tracks from the Dartmouth
database \cite{12}. These theoretical computations
can be used to take into account not only the general
abundances of heavy elements [Fe/H] in the stars, but
also the relative abundances of $\alpha$-process elements
[$\alpha$/Fe], with the helium abundance Y being proportional
to the metallicity. There is also the possibility
of specifying the enhanced helium abundances. The
evolutionary tracks for discrete masses were given in
[Fe/H] increments of 0.5~dex and [$\alpha$/Fe] increments
of 0.2~dex. We found the mass of each RR Lyrae star
by interpolating between all these parameters. To determine
[$\alpha$/Fe], we averaged the relative abundances
of magnesium, calcium, silicon, and titanium given
in \cite{1}. The abundances of all these four elements are
known for most stars in our sample. If there was no
information on one of the elements listed above, we
found the average for the other known $\alpha$-process elements
(the corresponding [$\alpha$/Fe] values are presented
in Table 1). There is only one star in our catalog with
no abundances of $\alpha$-process elements (UY CrB). In
order to find the mass of this variable, we used the
abundance [$\alpha$/Fe] = 0.0 typical of solar-metallicity
stars, similar to the metallicity of UY CrB 
(Increasing [$\alpha$/Fe] does not significantly 
change the mass of this star derived from the tracks.). The resulting
masses of the RR Lyrae stars are collected in
the last column of Table~1.

As an example, Fig.\,1a shows two evolutionary
tracks of different masses for the solar chemical
composition and two tracks for a metallicity a
factor of 100 lower, with typically increased relative
abundances of $\alpha$-process elements ([$\alpha$/Fe] = 0.4),
which are approximate upper and lower limits for the
RR Lyrae stars in our sample in the diagram. The
evolutionary tracks for lower masses and any metallicity
appear in the upper part of the diagram, as a
rule, at higher temperatures, while tracks for high
masses do not reach the instability strip in the lower
part of the diagram, and are completely outside its
low-temperature boundary. As a result, horizontal-branch
stars with the same atmospheric parameters
will have higher masses for lower metallicities. A decrease
in the relative $\alpha$-process element abundances
also results in a slight shift towards higher masses.
We can see from the log\,$T_{\textrm{eff}}$ -- log\,$g$ diagram that the
range of basic atmospheric parameters and the range
of masses are much narrower for high-metallicity
RR Lyrae stars, and the masses are, on average, lower
than for low-metallicity RR Lyrae stars.

Continuing our comparison with the absolute
magnitudes obtained via calibration, we plotted
the same evolutionary tracks in Fig.~1b. The
log\,$T_{\textrm{eff}}$ -- $M_{Ks}$ diagram shows that, if the absolute
magnitudes $M_{Ks}$ are used, the range of derived
masses for our RR Lyrae stars shifts considerably
towards higher masses, leaving the upper part of
the low-mass region of the diagram unoccupied, and
slightly displacing the lower boundary downwards.
As a result, the lowest RR Lyrae stars in this diagram
lie in the region that is not reached by the high-mass
evolutionary tracks from the theoretical computations
we used. That is, the absolute magnitudes of the
RR Lyrae stars found from their variation periods and
metallicities disagree with the theoretical positions
for horizontal-branch stars, making it impossible
to determine masses for many of them from these
data. The calibration of \cite{11} apparently requires some
correction.

Figure 2 shows the dependences of the derived
masses of the RR Lyrae stars on their atmospheric
parameters and chemical compositions. Different
symbols are used for stars belonging to different subsystems
of the Galaxy. We distinguished the subsystems
by applying the kinematic criterion from \cite{13},
where the components of the space velocities are used
to calculate probabilities for an RR Lyrae star to
belong to the thin disk, thick disk, or halo subsystems
(for more details, see \cite{1}). It was assumed here that
the the components of the stellar space velocities in
each of the subsystems have normal distributions.

Figure 2a displays the dependence of the mass on
the metallicity. The lower mass limit remains the
same for any metallicity, while the upper mass limit
grows linearly over the entire range of metallicities.
In the range $\rm{[Fe/H]} > -1.0$, however, we observe a
jump-like stronger concentration of stars toward the
lower mass limit than in the lower-metallicity range.
This behavior is completely unrelated to whether the
stars' kinematics ascribe them to a particular subsystem,
i.\,e., to their space velocities, but is related to
their metallicities: all metal-rich RR Lyrae stars in
the thick disk and halo have low masses. Note that
the transition through the boundary value, 
$\rm{[Fe/H]} = -1.0$, also results in an abrupt increase in the scatter
of the distances from the Galactic plane, as well as an
increase in the dispersion of the space velocities for
our RR Lyrae stars (see [\cite{1}, Fig. 1]). This behavior
of the velocity dispersion motivated the traditional
subdivision of RR Lyrae stars at this metallicity into
those belonging to the thick disk and halo, despite
the fact that it is precisely the kinematics that defines
the spatial distribution of the stars in the subsystems.
Nevertheless, we suggest that the jump in the masses
correlates with the jump in the velocities, simply because
both parameters are related through the metallicity,
although, physically speaking, the mass does
not depend on the velocity. Note that globular clusters
also exhibit abrupt changes in their spatial and kinematic
properties when the same metallicity is crossed
(e.g., \cite{14}).

The expected exponential relations are found for
each subsystem's RR Lyrae stars in the log\,$g$ -- $M/M_{\odot}$
diagram, Fig.~2b. The sequences of thin-disk and halo 
RR Lyrae stars essentially do not meet; as we
can see from the [Fe/H] -- $M/M_{\odot}$ diagram, the metallicities
of the stars in these subsystems do not overlap.
In addition, the masses of metal-rich RR Lyrae stars
are always less than those of metal-poor ones for
a given log\,$g$. On the other hand, the thick-disk strip
partially covers the sequences of both subsystems,
due to its very wide metallicity range. Further, the
log\,$T_{\textrm{eff}}$ -- $M/M_{\odot}$ diagram 
in Fig.~2c shows the masses
of RR Lyrae stars in different subsystems at different
temperatures. Here, we note that, independent of the
kinematic membership in a particular subsystem, the
upper mass limit for the RR Lyrae stars increases
with decreasing temperature, while the lower limit
is independent of temperature. The last diagram,
[$\alpha$/Fe] -- $M/M_{\odot}$ in Fig. 2d, also displays a mass
increase with increasing relative abundance of $\alpha$-
process elements. Since there are no RR Lyrae stars
with [$\alpha$/Fe]\,$\sim 0.15$, this suggests the presence of an
abrupt increase in the mean mass when this boundary
is crossed. This results from the fact that the relative
abundances of $\alpha$-process element for most thin-disk
RR Lyrae stars are nearly solar, but they abruptly
increase for metal-poorer stars.

\section {DISCUSSION} 

Our study shows that virtually all effective temperatures
($T_{\textrm{eff}}$) and surface gravities (log\,$g$) for field
RR Lyrae variables in the literature enter the instability
strip for corresponding theoretical evolutionary
tracks of horizontal-branch stars, enabling derivation
of the stellar masses. We suggest that this is
reflects the correctness of their derived atmospheric
parameters. On the other hand, there is no such
agreement for the absolute magnitudes $M_{Ks}$ derived
for these stars from calibration relations for variability
periods and metallicities. As a result, a large region
in the theoretically identified area of variable stars in
the log\,$T_{\textrm{eff}}$ -- $M_{Ks}$ diagram remains unoccupied, and
a considerable fraction of our RR Lyrae stars lie in a
region that they should not be able to enter according
to the theoretical computations.

Our checks show that the values of $M_{Ks}$ and
log\,$g$ are completely uncorrelated. This indicates
that the calibration of the absolute magnitudes of
the RR Lyrae stars from their variability periods and
metallicities is not quite correct. Therefore, we derived
the masses solely based on atmospheric parameters
determined using spectra. As a result, we
found that the lower mass limits for metal-rich and
metal-poor RR Lyrae stars coincide, and are approximately
half a solar mass. At the same time, the
upper mass limit decreases rapidly with increasing
metallicity. The low-metallicity, low-mass RR Lyrae
stars mainly belong to the halo subsystem, or less
frequently to the thick disk, indicating that they are
very old stars and pose no problem. However, in the
traditional assumption that a star’s mass loss in the
red-giant phase and the subsequent helium flash is of
order $(0.1-0.2)M_{\odot}$ (see \cite{6} and references therein),
so find that the initial masses for most of the metal-rich
stars are theoretically too low for them to reach the
horizontal branch during a time interval shorter than
the age of the thin disk subsystem.

An analysis of the elemental abundances in nearby
stars demonstrates that the thick disk also contains
old stars with the solar metallicity and low relative
abundances of $\alpha$-process elements, but with ages
exceeding 10 billion years (e.g., \cite{15}). However, the
RR Lyrae stars with solar abundances also exhibit
very ``young'' kinematics, typical of the thin disk
rather than this subsystem. If we suppose that these
are higher-mass stars --- Cepheids pulsating in an
overtone, i.\,e., with a shorter than usual period --- they
should also display systematically lower surface gravities
and temperatures characteristic of Cepheids, but
this is not observed.

There is another possible origin for these stars.
It was found for nearby stars displaying thin-disk
kinematics that a small number of stars with solar
abundances already appeared in the thin disk during
the very first stages of its formation (see [\cite{16}, Fig. 8]).
They are the so-called old metal-rich stars; it is believed
that they were born near the Galactic center
and then migrated from there due to perturbations
from asymmetric gravitational components, such as a
central bar or spiral density waves \cite{17}. According to
the Dartmouth evolutionary tracks, a star with an initial
mass of $1.05M_{\odot}$ and solar elemental abundances
should reach the horizontal branch in $\sim 10$ billion
years, which is usually believed to be the age of the
thin disk. However, in order to enter the instability
strip at the horizontal branch, stars with such high
masses must lose about half their mass during the
red-giant phase, as was suggested in \cite{6}.

We can further reduce the initial mass if the initial
helium abundances in the progenitors of metal-rich
RR Lyrae stars were higher. According to the Dartmouth
theoretical computations, even a star with a
mass of $0.8M_{\odot}$ and an initial helium abundance of
Y = 0.4 will reach the horizontal branch in $\sim9.3$ billion
years. In order to enter the instability strip,
such stars must lose $(0.2-0.3)M_{\odot}$ during the red-giant
phase and the helium flash, but this is a quite
realistic mass loss. In fact, modern computations
of stellar models \cite{18} demonstrate that stars with
this initial mass can lose even more than $0.2M_{\odot}$ via 
their stellar winds by the end of their evolution
on the red-giant branch, i.\,e., before the mass loss
after the helium flash. However, we must assume
the highest mass-loss rate in the empirical formula
describing this process in this case. Furthermore,
these same computations demonstrate that, with increasing
helium abundance, such stars arrive at the
higher-temperature part of the horizontal branch, like
our metal-rich RR Lyrae stars in Fig.~1. Stars with
enhanced helium abundances have already been detected
in the bulge. For example, the recent paper \cite{19}
presents helium-abundance estimates for the population
of RRab stars in the huge sample of RR Lyrae
stars from the OGLE IV survey \cite{20}, obtained based
on the fact that, according to the non-linear convective
pulsation model, the shortest period of the fundamental
RRab mode depends strongly on the helium
abundance in the star. In particular, it is concluded in
\cite{19} that their results cannot exclude the presence of
a small fraction of RRab stars with enhanced helium
abundances in the bulge, similar to the abundances
measured for bulge red-clump stars, with mean Y =
0.28-0.35 \cite{21}. Studies of the helium abundances
of field RR Lyrae variables that are currently in the
solar neighborhood would provide useful information
for verifying this conclusion.

\section*{ACKNOWLEDGMENTS}

The authors are grateful to the anonymous referee
for comments that have enabled us to improve
the paper. This work has been supported by the
Ministry of Education and Science of the Russian
Federation (state contracts 3.5602.2017/BCh and
3.858.2017/4.6).

\renewcommand{\refname}{REFERENCES}

\newpage

\begin{figure*}
\centering
\includegraphics[angle=0,width=0.99\textwidth,clip]{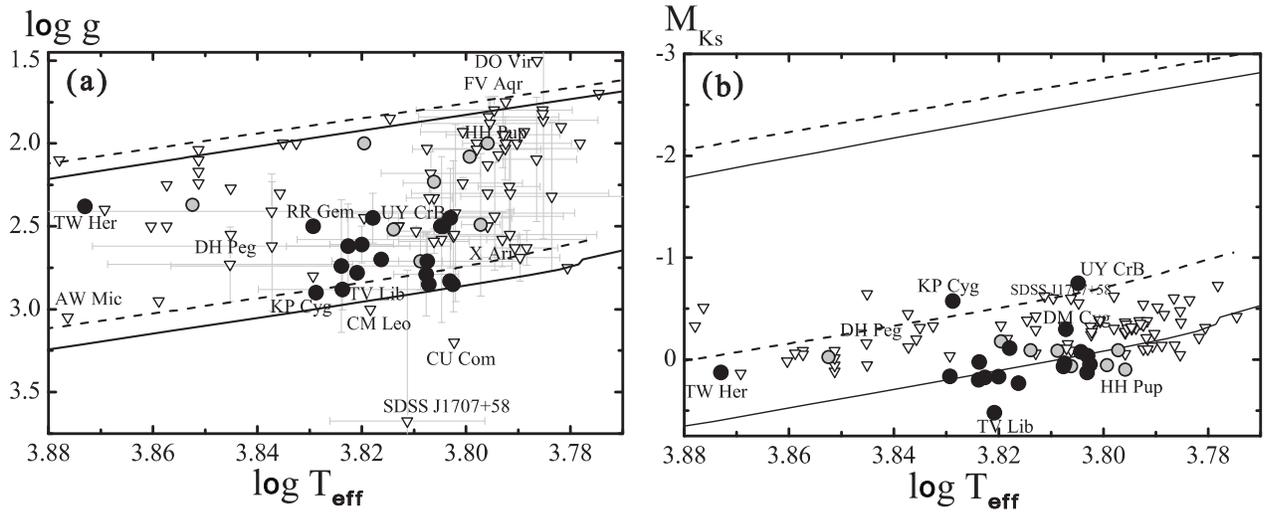}
\caption{Relations between the effective temperature, log\,$T_{eff}$, 
         and (a) surface gravity, log\,$g$, and (b) absolute 
         magnitude, $M_{Ks}$, for the RR Lyrae stars in our 
         sample. Black circles denote stars with 
         $\rm{[Fe/H]} > -0.5$, gray circles those with 
         $-0.5 > \rm{[Fe/H]} > -1.0$, and open triangles 
         the lowest-metallicity stars with $\rm{[Fe/H]} < -1.0$. 
         For RR Lyrae stars with atmospheric parameters determined 
         in multiple studies, the bars mark the highest and lowest 
         derived parameters for that RR Lyrae star. In both panels,
         two evolutionary tracks are plotted, for stars with the 
         solar chemical composition and masses of 0.49 and 0.54 
         $M_{\odot}$ (upper and lower solid curves, respectively). 
         The dashed curves show evolutionary tracks for stars with 
         masses of 0.52 and 0.75$M_{\odot}$, metallicities a 
         factor of hundred lower than the solar value, and relative 
         abundances of $\alpha$-process elements $\rm{[\alpha/Fe]} = 0.4$.
         The names of RR Lyrae stars that strongly deviate from 
         the highest-density concentration of data points in various 
         diagrams (including those in [1--3]) are marked.}
\label{fig1}
\end{figure*}

\newpage

\begin{figure*}
\centering
\includegraphics[angle=0,width=0.99\textwidth,clip]{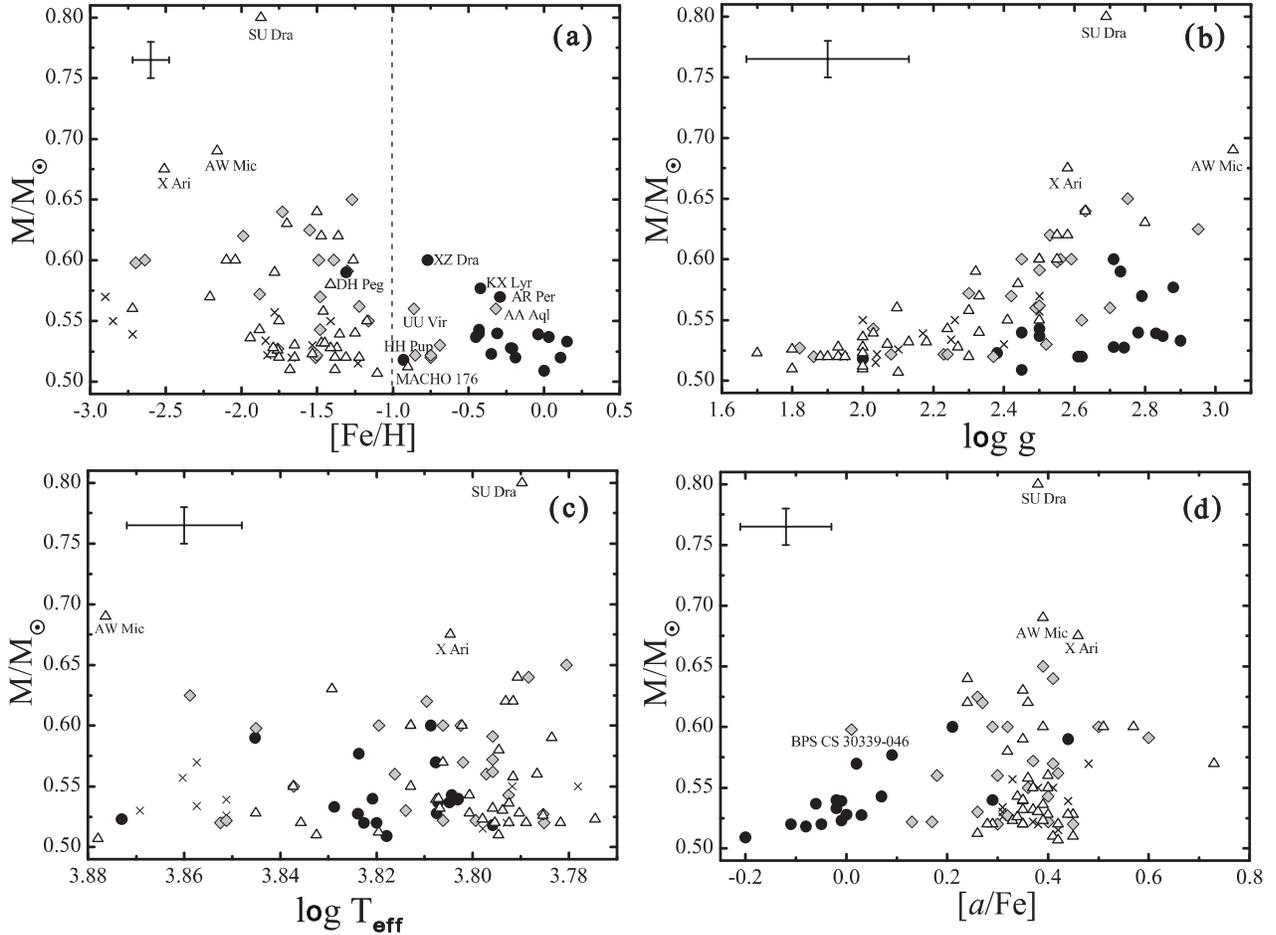}
\caption{Masses of RR Lyrae stars in our sample versus (a) 
         metallicity, (b) surface gravity, (c) effective 
         temperature, and (d) relative abundances of 
         $\alpha$-process elements. The black circles show 
         RR Lyrae stars belonging to the thin disk, according to
         their kinematics, gray diamonds those belonging 
         to the thick disk, open triangles those belonging to 
         the halo, and crosses RR Lyrae stars with no 
         specific classification. The vertical dashed 
         line is at $\rm{[Fe/H]} = -1.0$.}
\label{fig2}
\end{figure*}

\newpage
\clearpage

\newpage


\begin{table}[t!]

\caption{%
Astrophysical parameters of RR Lyrae stars}
\begin{center}
\begin{tabular}{c|c|c|c|c|c|c|c|c|c}
\hline \hline

\multicolumn{1}{c|}{\parbox{1.3cm}{Star}}&
\multicolumn{2}{c|}{\parbox{3.0cm}{Abundance, dex}}&
\multicolumn{3}{c|}{\parbox{2.0cm}{$T_{\textrm{eff}}$, K}}&
\multicolumn{3}{c|}{\parbox{2.0cm}{log\,$g$, dex}}&
\multicolumn{1}{c}{\parbox{1.2cm}{Mass, $M_{\odot}$}}\\
\cline{2-9}
&{[Fe/H]}&{[$\alpha$/Fe]}&{TP}&{min}&{max}&{TP}&{min}&{max}\\

\hline
   1  &  2     & 3      &  4   & 5   &  6   & 7    &  8   &9    &10   \\
\hline
SW And             & -0.22& 0.00& 6419& 6184& 6735& 2.71& 2.50& 2.85& 0.53\\
CI And             & -0.43& 0.07& 6373&     &     & 2.50&     &     & 0.54\\
DR And             & -1.37& 0.40& 6170& 6000& 6300& 2.00&     &     & 0.53\\
WY Ant             & -1.88& 0.34& 6319& 6150& 6487& 2.24& 2.20& 2.27& 0.54\\
XZ Aps             & -1.79& 0.45& 6319& 6200& 6438& 1.93& 1.90& 1.95& 0.53\\
BS Aps             & -1.48& 0.40& 6202& 6000& 6404& 2.03& 1.80& 2.18& 0.54\\
AA Aql             & -0.32& 0.18& 6550&     &     & 2.70&     &     & 0.56\\
SW Aqr             & -1.38& 0.28& 6200&     &     & 1.95&     &     & 0.52\\
BR Aqr             & -0.69& 0.26& 6515&     &     & 2.52&     &     & 0.53\\
DN Aqr             & -1.76& 0.34& 6100&     &     & 1.80&     &     & 0.53\\
FV Aqr             & -2.59& 0.41& 6200&     &     & 1.75&     &     &$<0.54$\\
X Ari              & -2.51& 0.46& 6378& 6109& 6950& 2.58& 2.10& 3.10& 0.68\\
ASAS J085254-0300.3& -1.53& 0.31& 7400&     &     & 2.40&     &     & 0.53\\
ASAS J162158+0244.5&-1.84 & 0.31& 7200&     &     & 2.25&     &     & 0.53\\
RS Boo             & -0.21& 0.03& 6666& 6233& 7275& 2.74& 2.40& 3.20& 0.53\\
ST Boo             & -1.73& 0.41& 6143& 6081& 6250& 2.63& 2.50& 2.71& 0.64\\
TW Boo             & -1.47& 0.37& 6250&     &     & 2.13&     &     & 0.53\\
BPS CS 22881-039   & -2.72& 0.40& 6117& 5950& 6170& 2.10& 1.85& 2.60& 0.56\\
BPS CS 22940-070   & -1.41& 0.38& 6191& 6130& 6300& 2.26& 1.85& 2.40& 0.55\\
BPS CS 30317-056   & -2.85& 0.41& 6000&     &     & 2.00&     &     & 0.55\\
BPS CS 30339-046   & -2.70& 0.01& 7000&     &     & 2.55&     &     & 0.60\\
W CVn              & -1.22& 0.42& 6250&     &     & 2.50&     &     & 0.56\\
UZ CVn             & -2.21& 0.73& 6400&     &     & 2.33&     &     & 0.57\\
YZ Cap             & -1.50& 0.40& 7100&     &     & 2.24&     &     & 0.52\\
RZ Cep             & -2.10& 0.57& 6500&     &     & 2.50&     &     & 0.60\\
RR Cet             & -1.48& 0.41& 6339& 5966& 6650& 2.42& 1.70& 2.77& 0.57\\
RX Cet             & -1.38& 0.45& 6800&     &     & 2.00&     &     & 0.51\\
UU Cet             & -1.36& 0.36& 6210& 6165& 6250& 2.58& 2.38& 2.71& 0.62\\
U Com              & -1.41& 0.44& 7000&     &     & 2.27&     &     & 0.53\\
CU Com             & -2.38& 0.38& 6343&     &     & 3.20&     &     &  -\\
UY CrB             & -0.45&     & 6380&     &     & 2.50&     &     & 0.54\\
W Crt              & -0.75& 0.17& 6400&     &     & 2.23&     &     & 0.52\\
XZ Cyg             & -1.50& 0.24& 6175&     &     & 2.63&     &     & 0.64\\
DM Cyg             &  0.03&-0.06& 6415&     &     & 2.85&     &     & 0.54\\
KP Cyg             &  0.15&-0.02& 6742&     &     & 2.90&     &     & 0.53\\
DX Del             & -0.31&-0.02& 6354& 6150& 6500& 2.45& 2.13& 2.73& 0.54\\
SU Dra             & -1.87& 0.38& 6161& 6083& 6300& 2.69& 2.50& 2.78& 0.80\\
SW Dra             & -1.27& 0.39& 6033&     &     & 2.75&     &     & 0.65\\
XZ Dra             & -0.77& 0.21& 6438& 6375& 6500& 2.71& 2.50& 2.85& 0.60\\
AE Dra             & -1.46& 0.51& 6525&     &     & 1.85&     &     & 0.51\\
SV Eri             & -1.99& 0.27& 6450& 6400& 6500& 2.53& 2.50& 2.55& 0.62\\
BK Eri             & -1.72& 0.36& 6840&     &     & 2.00&     &     & 0.51\\
CS Eri             & -1.70& 0.35& 6750&     &     & 2.80&     &     & 0.63\\

\hline

\end{tabular}
\end{center}
\end{table}

\newpage

\begin{table}[t!]
\begin{center}
\begin{tabular}{c|c|c|c|c|c|c|c|c|c}

\hline \hline

   1  &  2     & 3      &  4   & 5   &  6   & 7    &  8   &9    &10   \\

\hline

SX For             & -1.80& 0.33& 5950&     &     & 1.70&     &     & 0.52\\
RR Gem             &  0.01&-0.20& 6750&     &     & 2.50&     &     & 0.51\\
SZ Gem             & -1.65& 0.42& 6050&     &     & 1.90&     &     & 0.52\\
TY Gru             & -1.88& 0.37& 6250&     &     & 2.30&     &     & 0.57\\
BO Gru             & -1.83& 0.37& 7100&     &     & 2.04&     &     & 0.52\\
TW Her             & -0.35&-0.01& 7465&     &     & 2.38&     &     & 0.52\\
VX Her             & -1.46& 0.36& 6188& 5975& 6525& 2.30& 2.05& 2.72& 0.56\\
VZ Her             & -1.30& 0.60& 6250&     &     & 2.50&     &     & 0.59\\
DH Hya             & -1.53& 0.39& 6280&     &     & 2.00&     &     & 0.52\\
DT Hya             & -1.23& 0.42& 6280& 6100& 6460& 2.04& 2.00& 2.10& 0.52\\
V Ind              & -1.45& 0.35& 6409& 6267& 6550& 2.18& 2.03& 2.29& 0.53\\
RR Leo             & -1.39& 0.50& 6400& 6300& 6500& 2.59& 2.50& 2.65& 0.60\\
SS Leo             & -1.75& 0.37& 6875& 6100& 7650& 2.41& 2.05& 2.50& 0.55\\
ST Leo             & -1.31& 0.29& 6150&     &     & 1.93&     &     & 0.52\\
CM Leo             & -1.93& 0.42& 6582&     &     & 3.00&     &     &$>1.5$\\
TV Lib             & -0.43& 0.29& 6620&     &     & 2.78&     &     & 0.54\\
TT Lyn             & -1.47& 0.24& 6189& 6016& 6500& 2.55& 2.45& 2.75& 0.62\\
RR Lyr             & -1.49& 0.29& 6345& 6125& 6500& 2.56& 2.13& 3.04& 0.60\\
CN Lyr             & -0.04&-0.01& 6355&     &     & 2.83&     &     & 0.54\\
IO Lyr             & -1.35& 0.35& 6420&     &     & 2.03&     &     & 0.54\\
KX Lyr             & -0.42& 0.09& 6663& 6495& 7000& 2.88& 2.75& 3.00& 0.58\\
MACHO 176.18833.411& -0.90& 0.26& 6600&     &     & 2.00&     &     & 0.51\\
Z Mic              & -1.51& 0.45& 6098& 5950& 6246& 1.86& 0.60& 2.03& 0.52\\
AW Mic             & -2.16& 0.39& 7522&     &     & 3.05&     &     & 0.69\\
RV Oct             & -1.64& 0.46& 6247& 6050& 6443& 1.84& 1.70& 1.94& 0.51\\
UV Oct             & -1.75& 0.35& 6243& 6050& 6435& 1.88& 1.70& 2.00& 0.52\\
V 413 Oph          & -0.75& 0.30& 7120&     &     & 2.37&     &     & 0.52\\
V 445 Oph          &  0.11&-0.05& 6647& 6450& 6818& 2.62& 2.43& 2.94& 0.52\\
AO Peg             & -1.26& 0.39& 6342&     &     & 2.55&     &     & 0.60\\
AV Peg             & -0.19&-0.11& 6607& 6513& 6700& 2.61& 2.48& 2.70& 0.52\\
BH Peg             & -1.17& 0.40& 6500&     &     & 2.50&     &     & 0.55\\
DH Peg             & -1.31& 0.44& 7002& 6500& 7278& 2.73& 2.50& 2.95& 0.59\\
AR Per             & -0.29& 0.02& 6422& 6315& 6500& 2.79& 2.50& 3.00& 0.57\\
RU Psc             & -2.04& 0.51& 6500&     &     & 2.50&     &     & 0.60\\
HH Pup             & -0.93&-0.08& 6250&     &     & 2.00&     &     & 0.52\\
V 701 Pup          & -2.90& 0.48& 7200&     &     & 2.50&     &     & 0.57\\
VW Scl             & -1.22& 0.35& 6850&     &     & 2.30&     &     & 0.52\\
SDSS J170733.93+585059.7& -2.79& 0.89&6475& 6250&  6700& 3.68& 2.38&4.20&-\\
VY Ser             & -1.78& 0.35& 6075& 5900& 6400& 2.32& 1.85& 2.75& 0.59\\
AN Ser             &  0.00&-0.20& 6575& 6500& 6650& 2.45& 2.30& 2.60& 0.51\\
V 456 Ser          & -2.64& 0.32& 6600&     &     & 2.45&     &     & 0.60\\
T Sex              & -1.55& 0.26& 7225&     &     & 2.95&     &     & 0.63\\
V 440 Sgr          & -1.16& 0.36& 6874& 6269& 7400& 2.62& 2.15& 2.93& 0.55\\
V 1645 Sgr         & -1.94& 0.39& 6200&     &     & 2.00&     &     & 0.54\\
W Tuc              & -1.76& 0.32& 6100&     &     & 1.82&     &     & 0.53\\
BK Tuc             & -1.65& 0.38& 6220&     &     & 2.07&     &     & 0.53\\
TYC 4887-622-1     & -1.79& 0.31& 7100&     &     & 2.10&     &     & 0.53\\
TYC 6644-1306-1    & -1.78& 0.33& 7250&     &     & 2.50&     &     & 0.56\\

\hline

\end{tabular}
\end{center}

\end{table}

\newpage

\begin{table}[t!]
\begin{center}
\begin{tabular}{c|c|c|c|c|c|c|c|c|c}

\hline \hline

   1  &  2     & 3      &  4   & 5   &  6   & 7    &  8   &9    &10   \\

\hline

TYC 8776-1214-1    & -2.72& 0.44& 7100&     &     & 2.17&     &     & 0.54\\
RV UMa             & -1.25& 0.35& 6413& 6370& 6500& 2.33& 2.27& 2.50& 0.54\\
TU UMa             & -1.41& 0.32& 6231& 6116& 6500& 2.44& 2.10& 2.75& 0.58\\
CD Vel             & -1.67& 0.38& 6208& 6050& 6366& 1.95& 1.70& 2.10& 0.52\\
ST Vir             & -0.85& 0.13& 6300&     &     & 2.08&     &     & 0.52\\
UU Vir             & -0.86& 0.30& 6269& 6225& 6333& 2.49& 1.97& 2.83& 0.56\\
UV Vir             & -1.10& 0.42& 7550&     &     & 2.10&     &     & 0.51\\
AS Vir             & -1.68& 0.41& 6232& 6000& 6436& 1.80& 1.70& 1.87& 0.51\\
DO Vir             & -1.57& 0.33& 6115&     &     & 1.50&     &     &$<0.5$0\\

\hline

\end{tabular}
\end{center}

\end{table}


\end{document}